\documentclass[12pt]{article}
\usepackage{latexsym}
\usepackage{graphicx}
\usepackage{pstricks}
\usepackage{epsfig}
\usepackage{caption}
\usepackage{psfrag}
\usepackage{amsmath,amssymb,dsfont}
\newcommand{\be}{\begin{equation}}
\newcommand{\ee}{\end{equation}}
\newcommand{\bea}{\begin{eqnarray}}
\newcommand{\eea}{\end{eqnarray}}

\input epsf

\begin{document}

\begin{titlepage}

\begin{flushright}
CPT-P003-2013
\end{flushright}
\vspace*{1.5cm}
\begin{center}
{\Large \bf Narrow Tetraquarks at Large $N$}\\[3.0cm]

{\bf Marc Knecht, $^a$} and {\bf Santiago Peris, $^b$}\\[1cm]

$^a$
Centre de Physique Th\'{e}orique\\ CNRS/Aix-Marseille Univ./Univ. du Sud Toulon-Var (UMR 7332)\\
CNRS-Luminy Case 907, 13288 Marseille Cedex 9, France\\[5mm]
$^b$
Department of Physics\\ Universitat Aut{\`o}noma de Barcelona, 08193 Barcelona, Spain.\\[0.5cm]

\end{center}

\vspace*{1.0cm}

\begin{abstract}

Following a recent suggestion by Weinberg, we use the large-$N$ expansion in QCD to discuss the decay amplitudes of tetraquarks into
ordinary mesons as well as their mixing properties. We find that the flavor structure of the tetraquark is a crucial ingredient to
determine both this mixing as well as the decays. Although in some cases tetraquarks should be expected to be as narrow as ordinary
mesons, they may get to be even narrower, depending on this flavor structure.

\end{abstract}

PACS numbers: 11.15.Pg, 12.38.Lg, 13.25.Jx, 14.40.Rt

\end{titlepage}

In a recent paper, Weinberg \cite{Weinberg} has pointed out that tetraquark mesons (i.e. those formed by two quarks and two antiquarks),
if they survive the large-$N$ limit \cite{tHooft}, would have decay rates proportional to $1/N$  and would, therefore, be as narrow as ordinary mesons.
In this paper we would like to point out that the flavor structure in a tetraquark meson is crucial to determine the behavior of its decay
amplitude at large $N$ and, although Weinberg's result is generically true, there are tetraquark mesons for which there is an extra
suppression resulting in a decay width which goes as $1/N^2$. Complications like the effects of mixing between ordinary mesons and
tetraquarks need to be discussed in general, although they are not always an essential ingredient to reach this  conclusion.

\begin{figure}[bh]
\renewcommand{\captionfont}{\small \it}
\renewcommand{\captionlabelfont}{\small \it}
\centering
\includegraphics[width=4in]{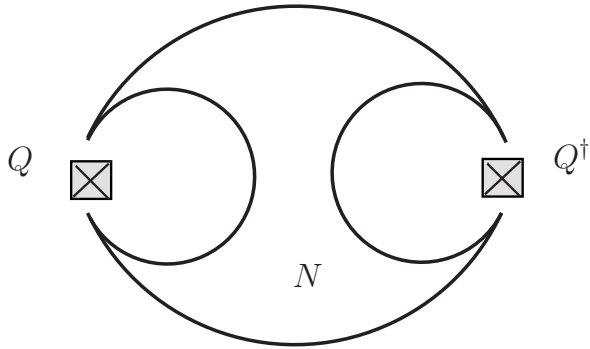}
\caption{Type-A diagram. In this figure, and all the ones that follow, it will be always understood that any number of gluon lines not
changing the large-$N$ behavior of the diagram should be added. This diagram is of order $N$, and this will be explicitly depicted with an
``$N$" in the diagram.}\label{f1}
\end{figure}

Let us start, following Ref. \cite{Weinberg}, by considering tetraquark interpolating fields as
\begin{equation}\label{tetraquark}
    Q(x)^{\alpha;\beta}_{AB;CD}=\ \mathbf{B}_{AB}^{\alpha}(x) \ \mathbf{B}_{CD}^{\beta}(x)
\end{equation}
where
\begin{equation}\label{bilinear}
    \mathbf{B}_{AB}^{\alpha}(x)=\sum_a \overline{q}_A^a(x) \Gamma^{\alpha} q_B^a(x)\ ,
\end{equation}
and $q_A^a(x)$ is the quark field with flavor $A$ and color $a$ (in $SU(N)$) and  $\Gamma^{\alpha}$ are colorless matrices containing
spin information. We will always assume that the flavor indices $A,B$ in these quark bilinears are different, so that the vacuum
expectation value $\langle \mathbf{B}_{AB}^{\alpha}(x)\rangle_0$ identically vanishes \cite{Coleman}. This simplifies the following 
discussion somewhat.
Since the flavor structure in the first bilinear is such that $A\neq B$, there are only three nontrivial possibilities for the other
bilinear $\overline{q}_C q_D$ (note that $C \neq D$ as well): either $C=B$, or $D=B$ or, finally, $A \neq B  \neq C \neq D$, i.e. all the
flavor indices are different\footnote{The cases $A=C$ and $A=D$ are analogous to the cases $B=D$ and $B=C$ (respectively)  and need not
be discussed separately.}. These three possibilities will determine the possible quark contractions in the relevant Green's function,
which in turn will determine its large-$N$ behavior.

\begin{figure}[t]
\renewcommand{\captionfont}{\small \it}
\renewcommand{\captionlabelfont}{\small \it}
\centering
\includegraphics[width=6in]{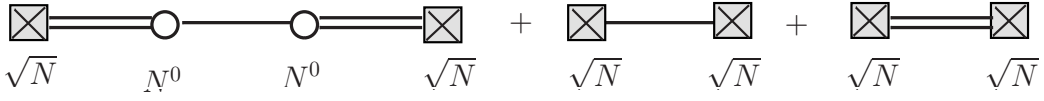}
\caption{Contribution of intermediate states to the type-A correlator of fig. \ref{f1}. A double line represents a tetraquark, a single line an
ordinary meson, a boxed cross signifies the insertion of the $Q$ operator, and an empty circle represents a vertex between a tetraquark and an
ordinary meson (which, in this case, implies mixing). The corresponding large-$N$ behavior for each of these terms in the diagram is explicitly
written.}\label{f1-states}
\end{figure}

Before entering the discussion of the different possible decay amplitudes for a tetra\-quark into ordinary mesons, it is of the utmost
importance to identify the tetraquark as a physical state in a Green's function, e.g. as a pole in the correlator\footnote{In the following,
we will suppress all the  indices in $Q(x)^{\alpha;\beta}_{AB;CD}$ for ease of notation unless required by clarity in the discussion.}
\begin{equation}\label{corr-Q}
    \langle Q(x) Q^{\dag}(0)\rangle_0\ .
\end{equation}
Therefore, as emphasized above, we now need to look at all the possible quark contractions in this correlator, and this is where the
flavor structure comes in.

\begin{figure}[hb]
\renewcommand{\captionfont}{\small \it}
\renewcommand{\captionlabelfont}{\small \it}
\centering
\includegraphics[width=4in]{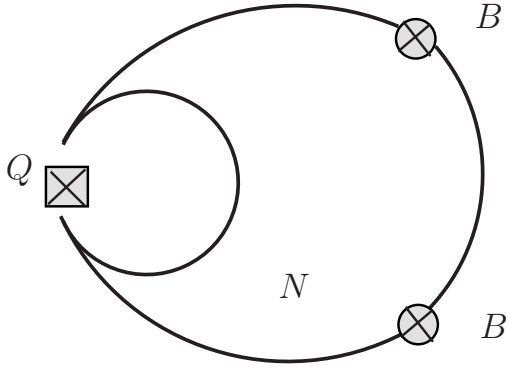}
\caption{Three-point function determining the decay of a type-A tetraquark into two ordinary mesons. A circled cross signifies the insertion
of the $\overline{q}q$ operator.}\label{f2}
\end{figure}

We will call ``type-A'' diagram a diagram like that in Fig. \ref{f1}, where the flavor in the tetraquark operator $Q(x)$ is of the
form $\overline{q}_A q_B \overline{q}_B q_C$ which allows an internal quark contraction between the two quarks with flavor index B.
This diagram is of order $N$. An example of this type of tetraquark is $\overline{u} d \overline{d}s$. Cutting vertically through this
diagram immediately reveals the presence of a potential four-quark intermediate state.

\begin{figure}[t]
\renewcommand{\captionfont}{\small \it}
\renewcommand{\captionlabelfont}{\small \it}
\centering
\includegraphics[width=3in]{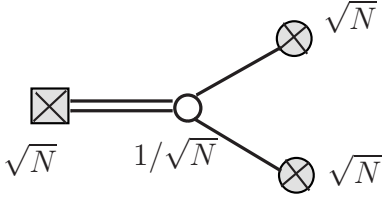}
\caption{Interpretation in terms of physical states of the contributions in fig. \ref{f2}. 
Symbols are like in fig. \ref{f1-states}.}\label{f2-states-a}
\end{figure}

Without knowing the solution to large-$N$ QCD it is not possible to know whether this four-quark intermediate state really exists and forms
the necessary pole in this Green's function. But if we assume that the state exists, large-$N$ can be used to predict how narrow it is and
how it mixes with ordinary $\overline{q}q$ mesons.  One can see in fig. \ref{f1-states} how to interpret the possible contributions to
this Green's function from all the possible intermediate states which are obtained by vertically cutting the diagram.
%
\begin{figure}[h]
\renewcommand{\captionfont}{\small \it}
\renewcommand{\captionlabelfont}{\small \it}
\centering
\includegraphics[width=3in]{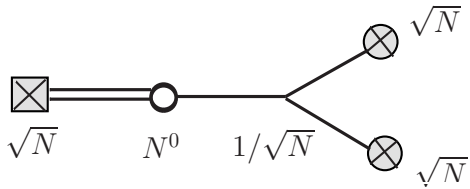}
\caption{Reinterpretation in terms of mixing between a tetraquark and an ordinary meson of the decay in fig. \ref{f2}.}\label{f2-states-b}
\end{figure}
%
This figure shows
the large-$N$ behavior of the different contributions, as determined  by matching the final value of $N$ obtained in  the diagram of
fig. \ref{f1}. For instance, looking at the last term in fig. \ref{f1-states} one finds  that the amplitude for the operator $Q$ to create
this tetraquark is of order $N^{1/2}$ since this amplitude appears squared. This matching has assumed that no cancellations between the
different contributions in fig. \ref{f1-states} takes place, as it is customarily done when following large-$N$ reasoning\footnote{An exception
to this rule is the mass of the $\eta'$ meson.\cite{Witten}}. Similarly, one also infers that the amplitude for $Q$ to create an ordinary meson
(second term in fig. \ref{f1-states}) is also of order $N^{1/2}$. Using these results, one finally obtains that the mixing between this
tetraquark and an ordinary meson (first contribution in fig. \ref{f1-states}) is of order $N^0$, and it is, therefore, not suppressed in the
large-$N$ limit. Because of this,  it may be very difficult to disentangle these tetraquarks from ordinary mesons, unless very precise
information on the position of the tetraquark pole is known. Furthermore, since this tetraquark is of the form
$\overline{q}_A q_B \overline{q}_B q_C$, with the flavor index $B$ contracted, it will fill out the same flavor representation as ordinary
$\overline{q}_A  q_C$ mesons, e.g. a whole nonet in flavour $SU(3)$.
%
%
\begin{figure}[t]
\renewcommand{\captionfont}{\small \it}
\renewcommand{\captionlabelfont}{\small \it}
\centering
\includegraphics[width=4in]{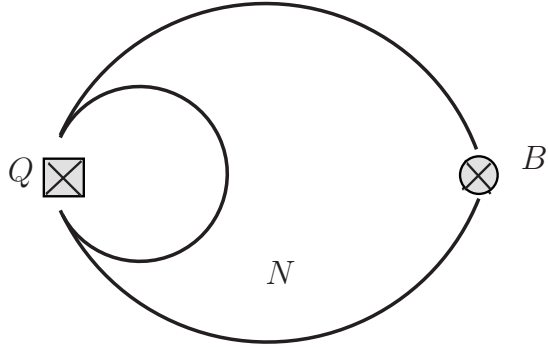}
\caption{Mixed correlator $\langle Q(x) \overline{q}(y)q(y) \rangle_0$.}\label{f3}
\end{figure}

In order to determine the width of a type-A tetraquark into two ordinary mesons, one may look at the three-point function depicted in
fig. \ref{f2} \footnote{The two $\overline{q}q$ mesons must of course have the right flavor structure to allow the contractions shown
in this figure.}

\begin{equation}\label{corr-QMM}
    \langle Q(x) \ \overline{q}(y)q(y)\  \overline{q}(z)q(z)\rangle_0\ ,
\end{equation}
which is of order $N$,  and look for the tetraquark pole. Its interpretation in terms of physical states is done in fig. \ref{f2-states-a}.
One concludes from this figure that the vertex between this tetraquark and two ordinary $\overline{q}q$  mesons is of order $N^{-1/2}$
(depicted as an empty circle in fig. \ref{f2-states-a}). Therefore, the decay width of a type-A tetraquark is of order $1/N$, i.e. as narrow as
an ordinary meson. This is the result obtained in ref. \cite{Weinberg}.

Fig. \ref{f2-states-b} helps one understand why a type-A tetraquark behaves like an ordinary meson in its decay:  one may reinterpret
this process as the mixing of the tetraquark with an ordinary meson (with an unsuppressed amplitude of order $N^0$), plus the subsequent decay
of the ordinary meson.
%
\begin{figure}[h]
\renewcommand{\captionfont}{\small \it}
\renewcommand{\captionlabelfont}{\small \it}
\centering
\includegraphics[width=6in]{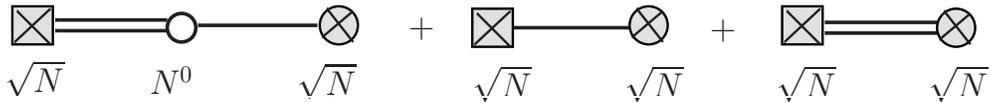}
\caption{Contributions from the intermediate physical states to the mixed correlator in fig. \ref{f3}.}\label{f3-states}
\end{figure}

The consistency of the conclusions drawn before may be cross-checked by looking at the mixed correlator in fig. \ref{f3}, which is or order
$N$, and its interpretation in terms of physical states, as is done in fig. \ref{f3-states}. From the first term in fig. \ref{f3-states}, one
again obtains that mixing between a tetraquark and an ordinary meson is of order $N^0$.

\begin{figure}[t]
\renewcommand{\captionfont}{\small \it}
\renewcommand{\captionlabelfont}{\small \it}
\centering
\includegraphics[width=4in]{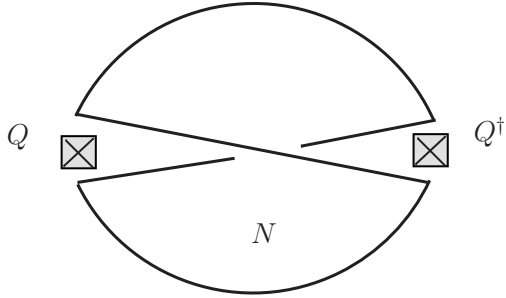}
\caption{Diagram for the two-point correlator of a tetraquark operator of type A$^{\prime}$ .}\label{f4}
\end{figure}

Tetraquarks with the flavor structure $\overline{q}_A q_B \overline{q}_C q_B$ (recall that $A\neq B$ and $C \neq B$),  do not allow the contractions of fig. \ref{f1}. An example of this tetraquark is $\overline{u}d \overline{s}d$. These tetraquarks will be called of ``type A$^{\prime}$ " because, after a simple rerun of the arguments presented for the previous diagrams in figs. \ref{f1}- \ref{f2-states-a}, the same final conclusion about the behavior of the width as for type-A tetraquarks  is reached, namely of order $1/N$. An important difference is, however, that these tetraquarks cannot mix with ordinary mesons because any intermediate state always contains 4 quarks, as fig. \ref{f4} shows. So, the first two contributions in fig. \ref{f1-states} are absent in this case.

\begin{figure}[hb]
\renewcommand{\captionfont}{\small \it}
\renewcommand{\captionlabelfont}{\small \it}
\centering
\includegraphics[width=4in]{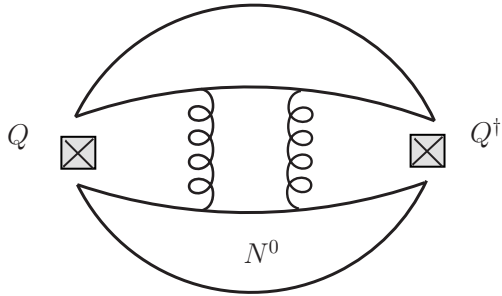}
\caption{Diagram for the two-point correlator of a tetraquark operator of type B. A minimum of two gluons are necessary to make this contribution connected.}\label{f5}
\end{figure}

We now come to the second type of  tetraquarks, which we will call ``type B",  i.e.  with all flavor indices different. This fact prevents
any internal quark contraction within the tetraquark operator $Q$. An example for this type of tetraquark is $\overline{u}d\overline{c}s$.
The corresponding diagram is depicted in fig. \ref{f5}. In order to make this contribution connected, a minimum number of two gluons is
necessary between the two quark loops, otherwise the diagram factorizes into two $\overline{q}q$ meson propagators which can contain no
tetraquark pole at leading order (see also below). Of course, in the diagram in fig. \ref{f5}, any number of gluons should be added as long as
they do not alter the large-$N$ counting, which is of order $N^0$. This diagram may also be present in the case of the type-A and A$^{\prime}$
tetraquarks, but its contribution remains subleading which is why we have not discussed it until now. When all the four flavor indices in $Q$
are different, the contribution in fig. \ref{f5} becomes leading and needs to be discussed separately.

\begin{figure}[t]
\renewcommand{\captionfont}{\small \it}
\renewcommand{\captionlabelfont}{\small \it}
\centering
\includegraphics[width=3in]{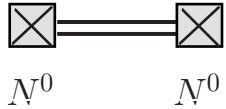}
\caption{Interpretation of the diagram in fig. \ref{f5} in terms of intermediate states.}\label{f5-states}
\end{figure}

\begin{figure}[hb]
\renewcommand{\captionfont}{\small \it}
\renewcommand{\captionlabelfont}{\small \it}
\centering
\includegraphics[width=4in]{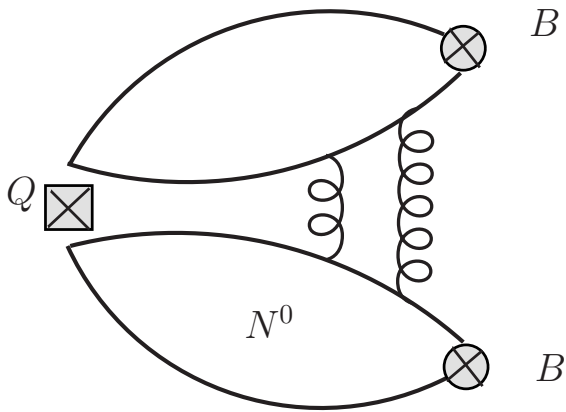}
\caption{Diagram for correlator governing the decay of a tetraquark operator of type B.}\label{f6}
\end{figure}

\begin{figure}[t]
\renewcommand{\captionfont}{\small \it}
\renewcommand{\captionlabelfont}{\small \it}
\centering
\includegraphics[width=3in]{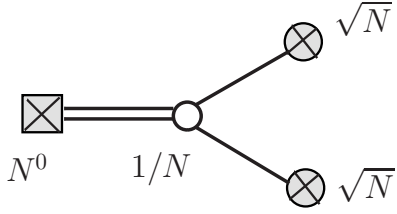}
\caption{Interpretation of the diagram in fig. \ref{f6} in terms of physical states.}\label{f6-states}
\end{figure}

Any intermediate state in fig. \ref{f5} always contains four quarks, which means that there is no mixing with ordinary mesons. Since the diagram is of order $N^0$, the amplitude for $Q$ to create one of these tetraquarks is also of order  $N^0$ (see fig. \ref{f5-states}). Fig. \ref{f6} shows the three-point function describing the decay of this tetraquark into ordinary mesons, which is of order $N^0$ as well. Its interpretation in terms of physical intermediate states in given in fig. \ref{f6-states}. One concludes therefore that the width for this type-B tetraquark is of order $1/N^2$ and is narrower than the type-A, and A$^{\prime}$  tetraquarks discussed previously. This concludes our discussion of the tetraquarks which may be excited from the vacuum by the action of the four-quark operator $Q$.
%

\begin{figure}[hb]
\renewcommand{\captionfont}{\small \it}
\renewcommand{\captionlabelfont}{\small \it}
\centering
\includegraphics[width=4in]{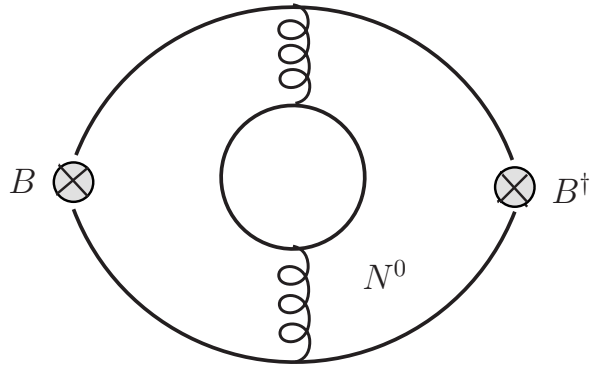}
\caption{Diagram showing the presence of a four quark state in the $\overline{q}q$ two-point correlator.}\label{f7}
\end{figure}

\begin{figure}[th]
\renewcommand{\captionfont}{\small \it}
\renewcommand{\captionlabelfont}{\small \it}
\centering
\includegraphics[width=6in]{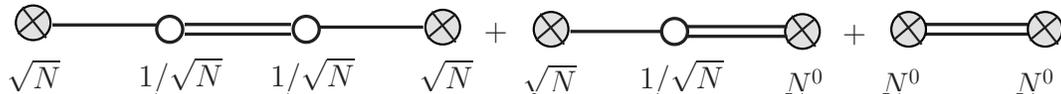}
\caption{Interpretation of the diagram in fig. \ref{f7} in terms of physical states.}\label{f7-states}
\end{figure}

We would like to point out, however, about the existence of another logical possibility. This consists of a $\overline{q}q$ bilinear
exciting a tetraquark meson through mixing, like in fig. \ref{f7}, which shows a diagram of order $N^0$. We will call these tetraquarks ``type
C". Clearly, a vertical cut of the diagram may contain a four quark intermediate state (Fig. \ref{f7-states} shows all the possible
intermediate states). Can this state be a tetraquark, i.e. can this singularity become a pole? Although it is true that the color flow in the
diagram shows that the intermediate state can be split in the product of two $\overline{q}q$ color singlets, this does not imply that these two
singlets necessarily have to become two separate meson states.  We think  it is not possible, on the sole basis of large-$N$ counting
arguments, to conclusively argue one way or the other without a more detailed dynamical knowledge of the QCD solution in this limit, but we see
no reason that would forbid the presence of a tetraquark pole in fig. \ref{f7}. If this is the case, these tetraquarks will have the flavor
structure of the type $\sum_B\overline{q}_A q_B \overline{q}_B q_C$ , i.e. they will also form a nonet, like type-A tetraquarks do.
%

\begin{figure}[h]
\renewcommand{\captionfont}{\small \it}
\renewcommand{\captionlabelfont}{\small \it}
\centering
\includegraphics[width=4in]{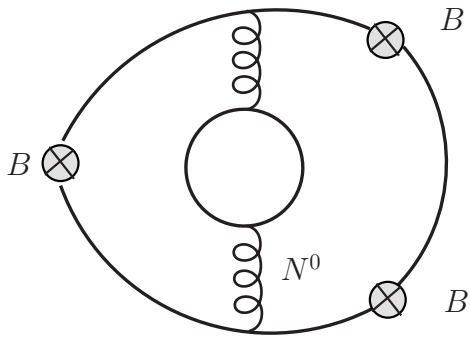}
\caption{Diagram depicting the three-point function governing the decay of a tetraquark of type C.}\label{f8}
\end{figure}

\begin{figure}[t]
\renewcommand{\captionfont}{\small \it}
\renewcommand{\captionlabelfont}{\small \it}
\centering
\includegraphics[width=4in]{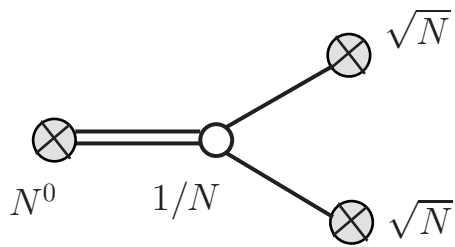}
\caption{Interpretation of the diagram in fig. \ref{f8} in terms of physical states.}\label{f8-states}
\end{figure}

As to the decay properties of a type-C tetraquark, one may look at the three-point function depicted in fig. \ref{f8}, which is of order $N^0$, and its corresponding interpretation in terms of physical states in fig. \ref{f8-states}. Since the vertex for this type of tetraquark to decay into ordinary mesons is of order $1/N$, its width will be of order $1/N^2$, i.e. as narrow as  type-B tetraquarks.
%

In conclusion, we have found that tetraquarks are narrow objects in the large-$N$ limit \cite{Weinberg}. They are at least as narrow
as ordinary mesons but they may get to be even narrower if the flavor structure is the appropriate one. There have been long discussions in the
literature trying to resolve the dichotomy between  the description in terms of a four-quark state, or a two-meson molecule \cite{Jaffe}. Our
conclusion is that, if the large-$N$ expansion is a good guidance, a tetraquark state should be narrow. If the state is broad, it is more
likely to be a two-meson bound state (molecule) resulting from an infinite chain of $1/N$-suppressed meson-meson interactions. This is
plausibly the way states like the $f_0(500)$ may be formed \cite{Pelaez}.
\vspace{-0.1in}
%

Perhaps the most promising tetraquark states from a theory point of view are those we called type B, with all the quark flavors different. For these tetraquarks the complications brought about by mixing will not be an issue since they cannot mix with ordinary mesons and, furthermore, they are expected to be long-lived as their width goes like $1/N^2$. It would be very interesting to see if the lattice could reveal their existence \cite{Wagner}. On the phenomenology side, there is mounting evidence about the existence of mesons with a four-quark content\cite{exp}. It remains to be seen whether the large-$N$ expansion can be helpful for explaining this phenomenology, and how it fares as compared to existing alternatives \cite{other}.

\vspace{1cm}

\textbf{Acknowledgements}

We would like to thank  M. Golterman and S. Weinberg for discussions.
This work has been supported by FPA2011-25948, SGR2009-894, the Spanish Consolider-Ingenio 2010 Program CPAN (CSD2007-00042).

\end{document}